\documentstyle[12pt,epsfig]{article}
\def\pipi{\mbox{$\pi^+\pi^-$}}
\def\mevc{\hbox{MeV\kern-2pt/\kern-2pt$c$}}
\def\gevc{\hbox{GeV\kern-2pt/\kern-2pt$c$}}

\begin{document}

\begin{center}
{\bf
THE EIKONAL APPROACH TO CALCULATION OF THE MULTIPHOTON EXCHANGE
CONTRIBUTIONS TO THE TOTAL CROSS SECTIONS OF \pipi{} ATOM INTERACTION WITH
ORDINARY ATOMS\\}
\vspace*{1cm}
L.AFANASYEV and O.VOSKRESENSKAYA\\
{\it Joint Institute for Nuclear Research, \\
141980 Dubna, Moscow Region,\\
Russia}
\end{center}

\vspace*{.5cm}
\begin{abstract}
{The total cross sections of the relativistic \pipi{} atom interactions
with ordinary atoms are obtained in the eikonal approach which takes into
consideration all multiphoton exchange processes. Contribution of these
processes strongly depends on the atom nucleus charge $Z$ and varies
from 1.5\% for Titanium ($Z=22$) to 14\% for Tantalum ($Z=73$)}

\end{abstract}

\vspace*{1cm}
An accurate measurement of the \pipi{} atom (dimesoatom) lifetime in the
experiment DIRAC \cite{dirac} will allow one to check a high precision
prediction of the Chiral Perturbation Theory. An interaction of the
$\pi^+\pi^-$ atoms with ordinary atoms is an essential part of this
experiment, as the atom observation bases on their breakup (ionization)
while passing through the target where they are produced.

Cross sections of $\pi^+\pi^-$ atoms with ordinary atoms were usually
calculated in the first Born approximation. In the paper \cite{tar91} the
cross sections for few low states of dimesoatom were calculated in
Coulomb-modified Glauber approximation. It has been shown that the
multiphoton exchange processes play a significant role in the interaction of
\pipi{} atoms with atoms.

In the Glauber approximation the total cross sections of the coherent
interaction of \pipi{} atom with ordinary atoms could be written \cite{tar91}
as:
\begin{equation}
\label{e1}
\sigma_{nlm}^{tot} = 2 \, \mbox{Re} \int d^2 b\: d^3 r
|\psi_{nlm}(\vec{r})|^2 \;\left[1-\exp{\left(i\chi(\vec{b}-\vec{s}/2) -
i\chi(\vec{b}+\vec{s}/2)\right)}\right] \,.
\end{equation}
Here $\vec s= \vec r_\perp$ is the projection of the vector $\vec r$ on the
plane of the impact parameter $\vec b$, $\psi_{nlm}(\vec{r})$
is the wave function of \pipi{} atom in the state with principal, orbital
and magnetic quantum numbers $n$, $l$ and $m$, respectively. The phase shift
$\chi(\vec b)$ is expressed via the screened Coulomb potential of the target
atom:
\begin{equation}
\label{e2}
\chi(\vec b)=\frac{1}{v} \int\limits_{-\infty}^{\infty}
U(\sqrt{b^2+z^2}\,)\:dz\,.
\end{equation}
Here $v$ is the velocity of dimesoatom in the lab frame.

Let us write (\ref{e1}) in another form taking into account the following
relations:
\begin{eqnarray}
\exp{(i\chi(\vec{b}))} & = & 1 -\gamma(b)  \label{e3}\, ,\\
\gamma(b) & = & \frac{1}{2\pi i} \int f(q) \exp{(-iqb)}\: d^2q \,,
\label{e4}\\
f(q) & = & \frac{i}{2\pi} \int \left[1- \exp{(i\chi(\vec{b}))} \right]
\,\exp{(i\vec{\vphantom{b} q}\vec{b})}\:  d^2b \nonumber\\
&=& i\: \int\limits_{0}^{\infty} \left[1- \exp{(i\chi(\vec{b}))} \right]
J_0(qb) b\: db
\, . \label{e5}
\end{eqnarray}
Here $f(q)$ is the amplitude of the elastic Coulomb $\pi A$-scattering
normalized by the relations:
\begin{eqnarray}
\sigma_{\pi A}^{tot} & = & 4\pi\: \mbox{Im}\, f(0) \label{e6}\, ,\\
\frac{d\,\sigma_{\pi A}}{d \vec q} & = & |f(q)|^2 \label{e7} \,.
\end{eqnarray}

Then it is easy to get the total cross section in the form
\begin{eqnarray}
\sigma_{nlm}^{tot} & = & 2 \int |f(q)|^2 \left[1-S_{nlm}(\vec q)\right]\:
d^2q \, ,\label{e8} \\
S_{nlm}(\vec q) & = &  \int |\psi_{nlm}(\vec r)|^2\: \exp{(i\vec q\vec r)}\:
d^3 r \label{e9} \,.
\end{eqnarray}
Here $S_{nlm}(\vec q)$ is the elastic form factor of the \pipi{} atom in the
state with the quantum numbers $n$, $l$ and $m$.

The expressions (\ref{e8}) and (\ref{e9}) together with the results
obtained in the paper \cite{af96} for the transition form factors of the
hydrogen-like atoms allow one to calculate the total cross sections for any
state of the dimesoatom. However, in this paper we only consider the cross
sections averaged over the magnet quantum number:
\begin{equation}
\label{e10}
\sigma_{nl}^{tot} =\frac{1}{2l+1} \sum_{m} \sigma_{nlm}^{tot} \, .
\end{equation}

The wave function of \pipi{} can be written as a product of the radial and
angular parts:
\begin{equation}
\label{e11}
\psi_{nlm}(\vec r) = R_{nl}(r)\: Y_{lm}(\Theta,\phi) \,.
\end{equation}
Taking into account the normalization
\begin{equation}
\label{e12}
\frac{1}{2l+1} \sum_{m} |Y_{lm}(\Theta,\phi)|^2 = 1 \,,
\end{equation}
we have
\begin{equation}
\sigma_{nl}^{tot}=4\pi\int |f(q)|^2 \left(1-S_{nl}(q)\right)\, q \,dq
\label{e14}\, ,
\end{equation}
\begin{eqnarray}
S_{nl}(q)&=&\frac{1}{2l+1}\sum_{m} S_{nlm}(q)=
\int |R_{nl}|^2 \exp{(i \vec q\vec r)} \: d^3 r=
4\pi \int\limits_{0}^{\infty}|R_{nl}|^2 \, \frac{\sin qr}{q} \,
r\, dr \nonumber \\ & = & \frac{4\pi}{q}\;
\mbox{Im}\int\limits_{0}^{\infty}|R_{nl}|^2 \exp{(i q r)}\, r\, dr \, .
\label{e13}
\end{eqnarray}

The radial wave function of the hydrogen-like atom is expressed in terms of
the Laguerre polynomial.  So integration in (13) is reduced to the
hypergeometric functions using the expression \cite{ryzh}
\begin{eqnarray}
\lefteqn
{\int\limits_{0}^{\infty}e^{-bx} \, x^\alpha \, L_{n}^{\alpha}(\lambda x) \,
L_{m}^{\alpha}(\mu x)=}
\nonumber \\
&&\frac{\Gamma(m+n+\alpha+1)}{\Gamma(m+1)\Gamma(n+1)}\:
\frac{(b-\lambda)^n(b-\mu)^m}{b^{n+m+\alpha+1}}
F\left[-m,-n,-m-n-\alpha; \frac{b(b-\lambda-\mu)}{(b-\mu)(b-\lambda)}\right]
\label{e15}
\end{eqnarray}

Finally for the form factor we have
\begin{equation}
\label{e16}
S_{nl}(q) = \frac{\sin{2n\phi}\;(\cos\phi)^{2l+4}}{n\;\sin{2\phi}}\;
F(l+1-n,l+1+n;1;\sin^2\phi)\, ,
\end{equation}
$$
\phi=\arctan{\left(\frac{nq}{\alpha m_\pi}\right)} \,.
$$

For the numerical calculation we use the Moli\'ere parametrization of the
Thomas-Fermi potential \cite{mol}
\begin{equation} \label{e17} U(r)=Z\alpha \sum_{i=1}^{3}
\frac{c_i e^{-\lambda_i r}}{r}\, ;
\end{equation}
which allows one to obtain the exact expression for the phase shift $\chi(b)$
\begin{equation}
\label{e18}
\chi(b)=\frac{2Z\alpha}{v}\sum_{i=1}^{3}c_i\; K_0(b\lambda_i)\, ,
\end{equation}
$$
c_1=0.35,\quad c_2=0.55,\quad c_3=0.1\, ;
$$
$$
\lambda_1=0.3\lambda_0,\quad \lambda_2=1.2\lambda_0,\quad \lambda_3=
6\lambda_0,
\qquad \lambda_0=m_e \alpha Z^{1/3}/0.885\, .
$$

The numerical results were calculated for the Titanium ($Z=22$) and Tantalum
($Z=73$) targets. The velocity of \pipi{} atom was taken as $v=1$.  Figures
\ref{f73cs}--\ref{f73l} show the most important dependencies.  We can
conclude that the contribution of the multiphoton exchange processes
strongly depends on the target atom nucleus charge $Z$ and varies from 1.5\%
for Titanium to 14\% for Tantalum.

In the DIRAC experiment the dimesoatom lifetime is going to be measured for
various targets with an accuracy of 10\% using the breakup probability
\cite{dirac} which is calculated base on the interaction cross sections. So
account of the multiphoton processes is essential for an interpretation
of the experiment results.

Authors would like to thank A.V.Tarasov from JINR, professor J.H\"ufner
from Heidelberg Universit\"at and professor B.Kopeliovich from
Max-Plank-Institut f\"ur Kernphysik for helpful discussions.
This work is partially supported by RFBR grant 97--02--17612.

\newlength{\pict}
\pict = 0.48\textwidth

\noindent
\parbox[t]{\pict}{
\mbox{\epsfig{file=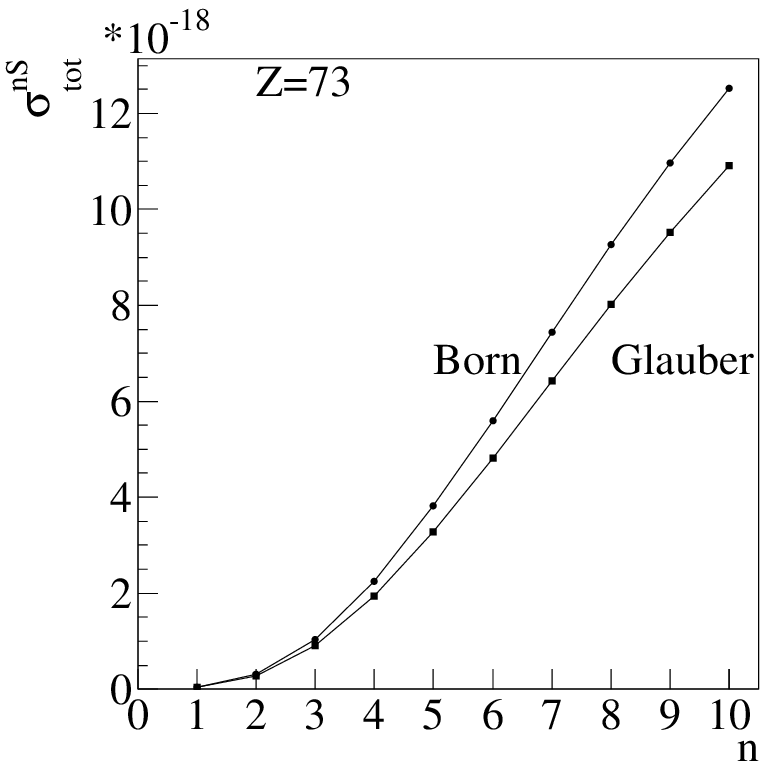,width=\pict}}

\refstepcounter{figure}
\label{f73cs}
Figure\enskip\ref{f73cs}: Total cross sections of \pipi-atom interactions
in $n$S states with Tantalum ($Z=73$) versus the principal quantum number
$n$ in the Born and Glauber approximation. }
\hfill \parbox[t]{\pict}{
\mbox{\epsfig{file=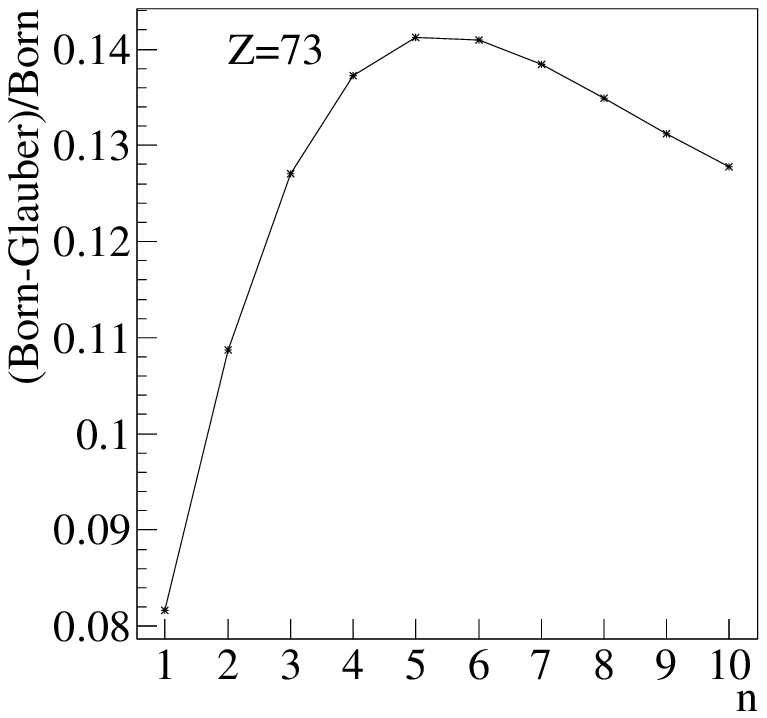,width=\pict}}

\refstepcounter{figure}
\label{f73r}
Figure\enskip\ref{f73r}: Relative difference between the total cross
sections of \pipi-atom interaction in $n$S states with Tantalum ($Z=73$)
calculated in the Born and Glauber approximation versus the principal
quantum number $n$.}

\noindent
\parbox[t]{\pict}{
\mbox{\epsfig{file=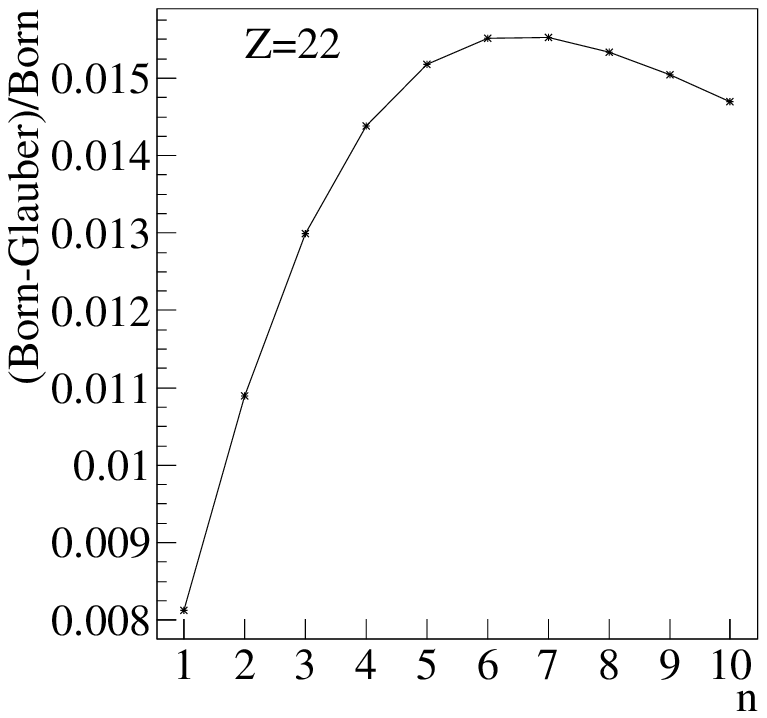,width=\pict}}

\refstepcounter{figure}
\label{f22r}
Figure\enskip\ref{f22r}: Relative difference between the total cross
sections of \pipi-atom interaction in $n$S states with Titanium ($Z=22$)
calculated in the Born and Glauber approximation versus the principal
quantum number $n$.}
\hfill \parbox[t]{\pict}{
\mbox{\epsfig{file=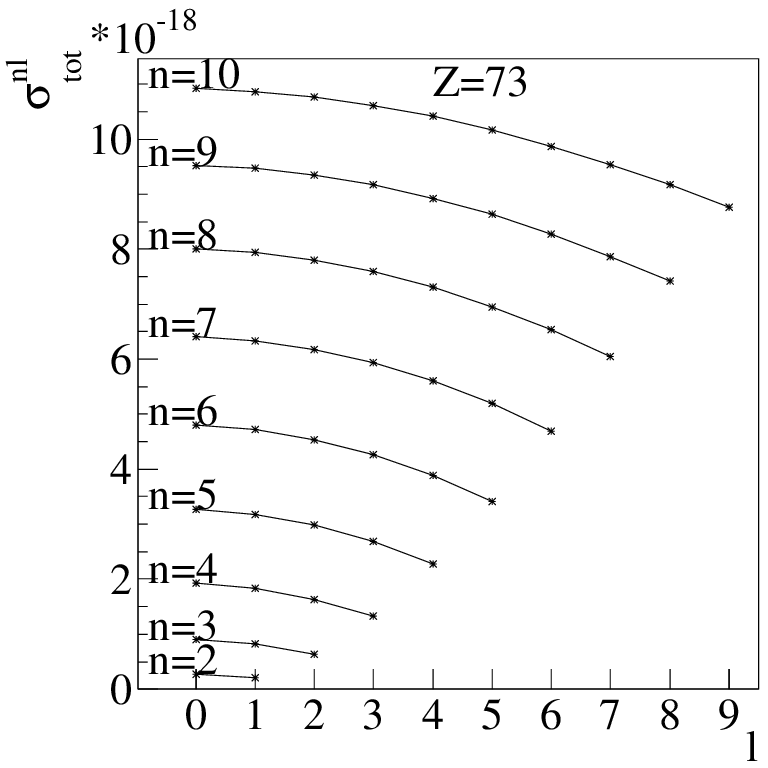,width=\pict}}

\refstepcounter{figure}
\label{f73l}
Figure\enskip\ref{f73l}: Dependence of the total cross sections of
\pipi-atom interaction with Tantalum on the orbital quantum number $l$ for
the various principal quantum number $n$.}

\end{document}